\documentclass[aps,prl,twocolumn,showpacs]{revtex4}
\usepackage{amssymb}
\usepackage{graphicx}
\usepackage[dvips]{epsfig}
\usepackage{dcolumn}
\usepackage{bm}
\usepackage{amsmath}
\begin{document}

\title{Spectral Polarization and Spectral Phase Control of Time-Energy Entangled Photons}
\author{Barak Dayan$^*$}
\author{Yaron Bromberg}
\author{Itai Afek}
\author{Yaron Silberberg}
\affiliation{Department of Physics of Complex Systems, Weizmann
Institute of Science, Rehovot 76100, Israel}

\begin{abstract}
We demonstrate a scheme to spectrally manipulate a collinear,
continuous stream of time and energy entangled photons to generate
beamlike, bandwidth-limited fluxes of polarization-entangled
photons with nearly-degenerate wavelengths. Utilizing an
ultrashort-pulse shaper to control the spectral phase and
polarization of the photon pairs, we tailor the shape of the
Hong-Ou-Mandel interference pattern, demonstrating the rules that
govern the dependence of this interference pattern on the spectral
phases of the photons. We then use the pulse shaper to generate
all four polarization Bell states. The singlet state generated by
this scheme forms a very robust decoherence-free subspace,
extremely suitable for long distance fiber-optics based quantum
communication.
\end{abstract}


\maketitle

Parametrically-generated polarization-entangled photons are a
primary resource in the fields of quantum communication and
quantum information \cite{QI}, motivating an ongoing search for
better means for the generation and control of high fluxes of such
photons \cite{HighFlux,kwiat:99:TwoType1,Kulik:01:Type_1_2}. In
particular, the generation of the singlet Bell-State has raised a
considerable interest due to the fact that it forms a
decoherence-free subspace (DFS), which is inherently immune
against to collective decoherence
\cite{kwiat:00:DFS,Brazilians:04}. The upper limit for the flux of
entangled photon pairs is typically set by the spatial divergence
of the down-converted photons, or by the repetition rate of the
pump pulses, in the case of pulsed down-conversion. In both cases,
the achievable flux is many orders of magnitude lower than the
physical upper limit, which is set by the down-converted bandwidth
\cite{Javanainen@Gould_PRA_1990,dayan:05:NonLinear}. Although
collinear and continuous down-conversion enables efficient
generation and collection of photon pairs, resulting in ultra-high
(bandwidth-limited) fluxes as high as $10^{12}\: s^{-1}$
\cite{dayan:05:NonLinear}, such single-mode configuration does not
readily enable polarization entanglement.

Pulse shaping techniques were recently introduced as a means to
control the spectral properties of broadband downconverted photons
\cite{pe'er:05:Shaping}.  In this work we use a phase and
polarization pulse-shaper \cite{brixner} to control both the phase
and the polarization of each of the spectral modes of
co-propagating entangled photons.
Our precise control of the phase and polarization in the frequency
domain is demonstrated by tailoring the shape of the
Hong-Ou-Mandel interference pattern \cite{Hong@Mandel_PRL_1987}.
Exploiting the fact that the photon pairs are hyperentangled, i.e.
entangled in more than one degree of freedom \cite{hyper}, we
treat the frequency domain as a two dimensional subspace and thus
use the pulse shaper as a Bell-State synthesizer, generating all 4
polarization Bell-states in a collinear, beam-like fashion, which
enables bandwidth-limited fluxes, and is suitable for fiber-optic
based quantum communication. The fact that in this scheme the
photons of each pair share the same single spatial mode and are
nearly degenerate, makes it a very robust and practical DFS.
Specifically, if the photons travel through the same fiber, this
singlet state is expected to be immune to geometric phase
\cite{ReviewQC,Banaszek04}, birefringence and all orders of
chromatic dispersion, preventing dephasing to distances which can
exceed 100 km (in typical optical fibers).\\

To illustrate the principles of our scheme, let us consider a
collinear, degenerate photon-pair at the state: $|\varphi\rangle =
\frac{1}{\sqrt{2}}\big(|2\rangle_H - |2\rangle_V\big)$ , where the
subscripts H,V indicate the horizontal and vertical polarizations
of the same spatial mode. This state can be generated by a
cascaded, collinear type-I down-conversion in two, orthogonal
crystals \cite{kwiat:99:TwoType1}. However, this is also the state
generated by a collinear type-II down-conversion in a single
crystal, which emits a pair of photons with orthogonal
polarizations X and Y, oriented at $45^\circ$ to H,V since
\cite{Kulik:01:Type_1_2}:
\begin{eqnarray} \label{Eq2}
|\varphi\rangle =\frac{1}{\sqrt{2}}\big(|2\rangle_H -
|2\rangle_V\big) = |1\rangle_X \: |1\rangle_Y \: .
\end{eqnarray}

We are specifically interested in the case where the
down-converted spectrum is significantly broader than the spectrum
of the pump, since this situation implies that the photons are
time and energy entangled, having a larger uncertainty in their
individual energies than in their collective energy (and the
opposite in the time-difference domain). This is typically the
case with continuous, degenerate spontaneous down-conversion,
especially in the configuration of two type-I crystals (as in our
experiment). The state of the photons in this case would therefore
be more accurately represented by:

\begin{eqnarray} \label{Eq3}
|\varphi\rangle = \int_0^{\omega_0} d\upsilon \:
\frac{g(\upsilon)}{\sqrt{2}}& \big( & |1 \rangle_{\omega_0 +
\upsilon,H} |1\rangle_{\omega_0 - \upsilon,H} - \nonumber\\ && |1
\rangle_{\omega_0 + \upsilon,V} |1\rangle_{\omega_0 - \upsilon,V}
\ \big) \: ,
\end{eqnarray}

\noindent where the spectral function $g(\upsilon)$ is determined
by the nonlinear coupling and the phase-matching conditions in the
crystal, and $\omega_0=\omega_p/2$, with $\omega_p$ being the pump
frequency. As is evident from Eq.~(\ref{Eq3}), although the two
photons share the same broadband spectrum, they are always on
opposite halves of the spectrum, having opposite detunings
$\pm\upsilon$ from $\omega_0$.
We therefore use the spectral degree of freedom to distinguish the
photons rather than the spatial modes. To clarify the polarization
entanglement of $|\varphi\rangle$ let us define two spectral modes
by assigning the subscripts $+,-$ to the spectral modes with
positive and negative detunings, respectively \cite{comment2}.
Adopting the notation $|H\rangle\equiv|1\rangle_H$, we may rewrite
Eq.~(\ref{Eq3}) as:
\begin{eqnarray} \label{Eq4}
|\varphi\rangle = \frac{1}{\sqrt{2}}\big( \ |H\rangle_+
|H\rangle_-  - |V\rangle_+ |V\rangle_- \: \big) \: .
\end{eqnarray}

\begin{figure}[tb]
\includegraphics[width=7.6 cm]{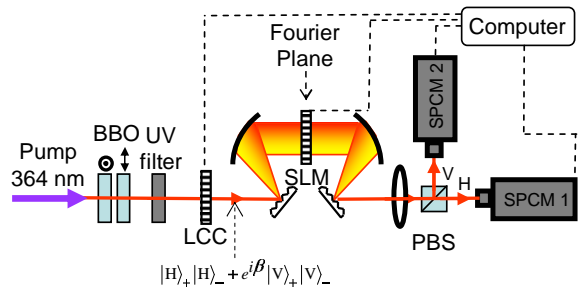}
\caption{\label{setup} (Color online) The experimental setup. In
addition to an initial harmonic separator filter, and final
bandpass filters ($70\:nm$ around $728\:nm$) in front of the
SPCMs, gold reflection coatings were used throughout the setup to
attenuate the UV pump with respect to the down-converted photons.}
\end{figure}

Thus, we see that the state $|\varphi\rangle$ is \textit{already}
a polarization Bell state, specifically the state
$|\phi^{(-)}\rangle_\textsc{h,v}$. This Bell state can be
transformed into the Bell state $|\psi^{(+)}\rangle_\textsc{h,v}$
by a simple rotation of $45^\circ$
, since $|\phi^{(-)}\rangle_\textsc{h,v}$ at the axes $H,V$ is
$|\psi^{(+)}\rangle_\textsc{x,y}$ at the axes $X,Y$. Similarly,
the Bell state $|\phi^{(+)}\rangle_\textsc{h,v}$ can be created
from $|\phi^{(-)}\rangle_\textsc{h,v}$ by introducing
polarization-dependent phase shift (for example, with a
birefringent material). However, the singlet state
$|\psi^{(-)}\rangle_\textsc{h,v}$ is unique. Having a zero
collective 'spin', the singlet state forms a DFS, and is not
affected by any mechanism that acts equally on both modes;
accordingly, it can not be created by such mechanisms. Since in
our case the two modes are the two halves of the down-converted
spectrum, in order to create the singlet state one must apply
\textit{spectrally-dependent} birefringence. Conveniently enough,
this is exactly what a pulse-shaper does. A pulse-shaper (see Fig.
\ref{setup}) separates the spectral components of the incoming
beam by a grating, and focuses them on a spatial-light modulator
(SLM), located at the Fourier-plane \cite{weiner:shaper}. The SLM
is composed of an array of liquid-crystal cells that induce a
voltage-controlled birefringence, i.e. a relative phase between
the two polarizations of the incoming light. A second set of a
curved-mirror and a reflection-grating performs the inverse
Fourier-transform and restores the collinear propagation of the
beam. Utilizing a pulse shaper, one can apply a phase-shift of
$\pi$ between the $X$ and $Y$ polarizations \textit{only on the
upper half of the spectrum} (a spectral '$\pi$ step-function').
This will create a $\pi$ phase shift between the $|X\rangle_+
|Y\rangle_-$ and the $|Y\rangle_+ |X\rangle_-$ components of the
$|\psi^{(+)}\rangle_\textsc{x,y}$ state, turning it into the
desired singlet state $|\psi^{(-)}\rangle_\textsc{x,y}$ (note that
once the singlet state is created, it remains the singlet state in
every polarizations basis, so
$|\psi^{(-)}\rangle_\textsc{x,y}=|\psi^{(-)}\rangle_\textsc{h,v}$).\\

Our experimental setup is depicted in Fig. \ref{setup}. The photon
pairs were generated by two adjacent, $1mm$ thick beta barium
borate (BBO) crystals, pumped with a continuous argon laser
($180\:mW$ at $363.8\:nm$, polarized along the $X$ axis) and
oriented to obtain collinear type-I down-conversion. By rotating
the pump polarization, and by making sure the entire optical setup
had the same attenuation for all the polarizations, we ensured
that both the generation and the detection of the photon pairs
occurred at the same probability for $V$ and $H$ polarizations. A
computer controlled liquid crystal cell (LCC), placed immediately
after the two crystals, controlled the relative phase $\beta$
between the $H$ polarized pairs and the $V$ polarized ones. The
photons were then directed to a phase and polarization
pulse-shaper \cite{brixner}, which included two orthogonal SLMs,
one applying a spectral phase $\theta_\textsc{x}(\omega)$ to $X$
polarized photons, and the second applying a spectral phase
$\theta_\textsc{y}(\omega)$ to $Y$ polarized photons. The
out-coming beam was then split by a polarizing beam-splitter (PBS)
to $H$ and $V$ polarized beams, which were focused onto two
single-photon counting modules (SPCMs). The coincidence counts
were
recorded with a temporal resolution of $12.5\:ns$.\\

\begin{figure} [t!b]
\includegraphics[width=7.6cm]{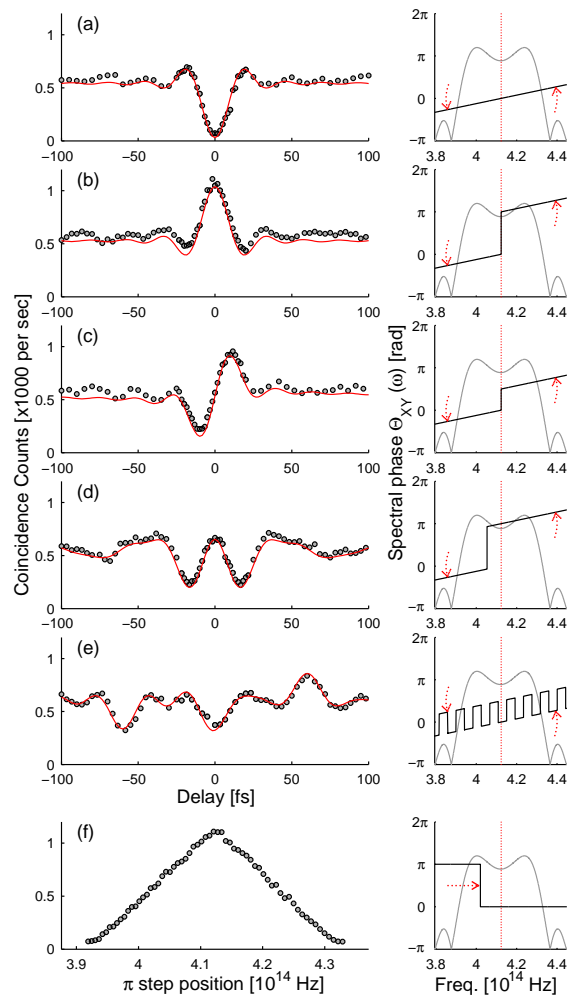}
\caption{\label{shaping} (Color online) (a)-(e): The left column
shows the experimental results (circles) and theoretical
calculation (line) of the coincidence count $R_c$ vs. the applied
delay between the $X$ and $Y$ polarizations, for various spectral
phase functions $\Theta_\textsc{xy}(\omega)$, which are depicted
in the right column (black), together with the down-converted
spectrum (gray). (f): The same as (b), except that the delay is
set to zero and the $\pi$-step position is scanned over the
spectrum.
The power spectrum used in the calculation of
Eq.~(\ref{Rc2}) was obtained by fitting the curve (f) to the
integral of the power spectrum expected for type-I
down-conversion. No other fitting procedures were applied.}
\end{figure}

A key point in this scheme is that the PBS at the
output of the pulse shaper mixes the $X,Y$ polarizations, and thus
induces a quantum-interference between the two two-photon
wave-functions which contribute to coincidence detections in the
SPCMs. Using the LCC, we set the phase $\beta$ to $\pi$, thus
creating the state of Eq.~(\ref{Eq3}) at the input of the pulse
shaper, in which each SLM affected only one photon of the pair.
This allowed us to induce a relative delay between the photons by
simply applying linear spectral phases with opposite slopes on the
two SLMs: $\theta_\textsc{x,y}(\omega)=\pm\tau\omega$
\cite{pe'er:05:Shaping}. This configuration is completely
equivalent to the two-photon interference experiment by Hong, Ou
and Mandel (HOM) \cite{Hong@Mandel_PRL_1987}, and indeed, a scan
of the relative delay between the photons reproduced the so-called
'HOM dip' in the coincidence counts with a visibility of $0.79\pm
0.01$ (Fig.~\ref{shaping}a).

Since our setup enables the application of arbitrary phases to
each of the spectral and polarization modes of the photons, it
allows convenient exploration of the \textit{spectrally-dependent}
quantum interference that governs the HOM interference pattern,
which was previously explored only by utilizing interference
filters or spectrally-independent phase shifters
\cite{PreviousHOM}. In pulsed down-conversion, or with photons
from independent single-photon sources, the HOM interference
pattern depends on the overlap between the (independently defined)
temporal envelopes of the two photons. This is not the case with
time and energy entangled photons, which can exhibit a zero
coincidence rate even if the photons do not overlap temporally at
the beam splitter \cite{Pittman@Shih_PRL_1996}. At low photon
fluxes, the coincidence rate, $R_c$, between the two output ports
of the HOM interferometer is given by: \cite{MW95}
\begin{eqnarray} \label{Rc1}
R_c \propto \int_T d\tau\
|\langle0|\hat{E}^{+}(\mathbf{r_2},t+\tau)\hat{E}^{+}(\mathbf{r_1},t)|\varphi\rangle|^2
\ .
\end{eqnarray}
 Where $\mathbf{r_1}$ and $\mathbf{r_2}$ are the
detectors locations, $\tau$ is the time difference between the two
detection events, and $T$ is the coincidence-circuit temporal
resolution. Inserting $|\varphi\rangle$ from Eq.~(\ref{Eq3}) into
Eq.~(\ref{Rc1}) and accounting for the spectral phases applied by
the SLMs, one gets the following expression for $R_c$:
\begin{eqnarray}\label{Rc2}
R_c  \propto  \Bigl[ 1-\int_0^{\omega_0}  d\upsilon
   \: \big|g(\upsilon)\big|^2
   e^{\imath(\Theta_\textsc{xy}(\upsilon)-\Theta_\textsc{xy}(-\upsilon))}
   \:  \Bigr]\ ,
\end{eqnarray}
\noindent where
\begin{eqnarray} \label{t=tx+ty}
\Theta_\textsc{xy}(\upsilon)\equiv
\theta_\textsc{x}(\omega_0+\upsilon)-\theta_\textsc{y}(\omega_0+\upsilon)
\ .
\end{eqnarray}

As evident from Eq.~(\ref{t=tx+ty}), only the spectral phase
\textit{difference} between the interferometer arms counts. Thus,
when identical phase functions were applied by both SLMs, the
delay scan yielded the same HOM dip as depicted in
Fig.~\ref{shaping}a. For this reason, in  Fig.~\ref{shaping} we
presented our results as they are related to the spectral phase
difference between the SLMs  $\Theta_\textsc{xy}(\omega)$.
Moreover, Eq.~(\ref{Rc2}) indicates that only spectral phase
difference that is antisymmetric with respect to $\omega_0$
affects the two-photon interference pattern \cite{Walmsley97}.
This effect of nonlocal dispersion-cancellation, which was
previously predicted \cite{Franson:92:DisCanc,SKC92:PRA} and
demonstrated \cite{SKC92:PRL} for first-order dispersion, was
demonstrated in our setup for \textit{all odd orders} of
dispersion by the fact that applying arbitrary spectral phases
whose difference was symmetric about $\omega_0$ had no effect on
the HOM dip.  Another interesting result of Eq.~(\ref{Rc2}) is
that the coincidence function $Rc$ could be zero only at one value
of the delay between the interferometer arms; such a zero occurs
if and only if there is no antisymmetric spectral phase difference
between the interferometer arms, and then the shape of the HOM dip
is dictated \textit{only by the power spectrum} of the
down-converted photons.

Figures \ref{shaping}b-\ref{shaping}e demonstrate our ability to
arbitrary tailor the shape of the HOM interference pattern, with
excellent agreement with the theoretical calculation according to
Eq.~(\ref{Rc2}). It is interesting to note the difference between
Fig. \ref{shaping}a and \ref{shaping}b at zero delay. The first
occurs when the state of the photon pairs is essentially the
"untouched" initial $|\phi^{(-)}\rangle_\textsc{h,v}$, which is
composed only of two horizontal photons or two vertical ones, and
therefore yields zero coincidences at the two ports of the PBS.
The second occurs when a $\pi$ step-function is applied by one of
the SLMs, with the step located at the frequency $\omega_0$. In
this case the state of the pairs becomes the singlet state
$|\psi^{(-)}\rangle_\textsc{h,v}$, which always results in
coincidences between the output ports of the PBS. Since the
difference between the two states results from a destructive
spectral interference which turns into a constructive one due to
the presence of the spectral $\pi$ step-function, we can induce
this change gradually by scanning the position of the
step-function along the down-converted spectrum. The result of
such a scan appears in Fig. \ref{shaping}f. Since the coincidence
rate is proportional to the relative part of the spectrum that
experiences the phase-flip, the actual down-converted
power-spectrum is the absolute value of the derivative of this
graph. The power spectrum derived from Fig.~\ref{shaping}f was
used in the calculation of the theoretical curves in
Figs.~\ref{shaping}a-\ref{shaping}e.\\
\begin{figure}[t!b]
\includegraphics[width=7.6cm]{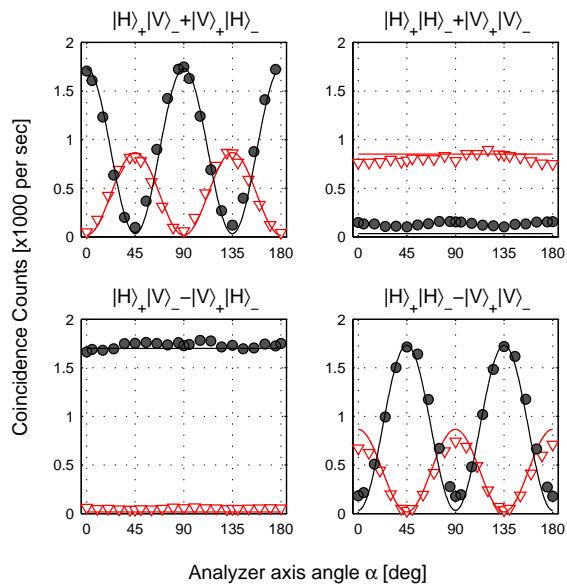}
\caption{  (Color online) Coincidence measurements illustrating
the production of all $4$ Bell states in our system. The plots
show the experimental coincidence rates $R_{\alpha,\alpha^\bot}$
(circles) and $R_{\alpha,\alpha}$ (triangles) vs. the angle of the
analysis axis $\alpha$ with respect to $H$, as compared to the
theoretical calculations (line) which assume pure Bell states.
$R_{\alpha,\alpha^\bot}$ indicates coincidence of orthogonally
polarized photons (at polarizations $\alpha,\alpha^\bot$), and
$R_{\alpha,\alpha}$ indicates coincidences of photons which are
both $\alpha$ polarized. The visibilities of the
$|\psi^{(+)}\rangle_\textsc{h,v}$
($|\phi^{(-)}\rangle_\textsc{h,v}$) state are $0.90\pm0.02$ $(0.93
\pm 0.01)$ for the $R_{\alpha,\alpha}$ curve and $0.90\pm0.01$
 $(0.81\pm0.02)$ for the $R_{\alpha,\alpha^\bot}$ curve.\label{fig:bell} }
\end{figure}

Finally, we used the computer-controlled pulse shaper and LCC to
generate the four polarization Bell-states. By setting the phase
$\beta$ to $\pi$ or $0$ we generated the states
$|\phi^{(-)}\rangle_\textsc{h,v}$ and
$|\phi^{(+)}\rangle_\textsc{h,v}$, respectively. The remaining two
Bell states were generated from these two states by applying the
spectral $\pi$ step-function, located at $\omega_0$ on one of the
SLMs. As explained earlier, since at the principal axes $X,Y$ of
the SLMs, $|\phi^{(-)}\rangle_\textsc{h,v}$ is
$|\psi^{(+)}\rangle_\textsc{x,y}$, it is turned into
$|\psi^{(-)}\rangle_\textsc{x,y}=|\psi^{(-)}\rangle_\textsc{h,v}$
by such a phase step-function. Similarly,
$|\phi^{(+)}\rangle_\textsc{h,v}=|\phi^{(+)}\rangle_\textsc{x,y}$
is turned into
$|\phi^{(-)}\rangle_\textsc{x,y}=|\psi^{(+)}\rangle_\textsc{h,v}$
by the same spectral phase function. In order to measure the
generated Bell states we introduced a half-wave plate between the
output port of the pulse shaper and the PBS, with its principal
axis rotated by $\alpha/2$ from the axis $H$
($0\leq\alpha\leq180^\circ$). Thus, we measured for each state the
coincidence counts rate $R_{\alpha,\alpha^\bot}$ at orthogonal
polarizations $\alpha,\alpha^\bot$ which were rotated by the angle
$\alpha$ with respect to $H,V$. Similarly, we measured the
coincidence rate of two photons at the same polarization axis,
namely, $R_{\alpha,\alpha}$. We did so by locating a polarizer at
the angle $\alpha$ at the output port of the pulse shaper, and
using the following half-wave plate to rotate this polarization to
the axis $X$. By doing this, $\alpha$ polarized photons had an
equal, independent probability to be directed to either of the
SPCMs, which meant that the measured coincidence rate corresponded
to half of the actual coincidence rate of $\alpha$ polarized
photons. The experimental results are depicted in Fig.
\ref{fig:bell}, together with the theoretical curves.
This set of measurements provides sufficient information about the
density matrix of the generated states to deduce lower bounds on
their fidelities with the desired Bell states, indicating that the
Bell states $|\psi^{(-)}\rangle_\textsc{h,v},
|\psi^{(+)}\rangle_\textsc{h,v}, |\phi^{(-)}\rangle_\textsc{h,v},
|\phi^{(+)}\rangle_\textsc{h,v}$, were produced with fidelities
which are equal to or grater than $0.90\pm0.01$, $0.88\pm0.01$,
$0.85\pm0.02$, $0.84\pm0.01$, respectively. Let us also note that
three out of the four generated states
($|\psi^{(-)}\rangle_\textsc{h,v},
|\psi^{(+)}\rangle_\textsc{h,v}$, and
$|\phi^{(-)}\rangle_\textsc{h,v}$) demonstrate non-classical
behavior by violating the Cauchy-Schwartz inequality for the
second order correlation functions, with
$R_{\alpha,\alpha}R_{\alpha^\bot,\alpha^\bot}<R_{\alpha,\alpha^\bot}^2$.\\

The ability to generate Bell-states in a collinear fashion which
allows coupling into an optical fiber bears great significance for
quantum communication (see, for example,~\cite{Banaszek04}). We
believe that our scheme will allow efficient coupling of ultrahigh
fluxes of entangled pairs into fibers, fluxes which are many
orders higher than those achievable by pulsed systems
\cite{dayan:05:NonLinear}. Note that our scheme also enables the
assignment of different Bell-states to different spectral
mode-pairs, thus creating a quantum equivalent to
wavelength-division multiplexing (WDM) in classical optical
communication systems. Additionally, the singlet state of this
scheme, which appears to be the most robust DFS that can be
supported in a single-mode fiber, could be used to conveniently
transport indistinguishable photon pairs over large distances on
the same fiber, and then separate them deterministically. This is
due to the fact that any PBS located at the output of the fiber
will separate the photons, one to each port, with no need to
compensate for the (usually time-dependent) birefringence. Note
that in this case the photons share the same spectrum, and are
time and energy entangled. These abilities could become useful in
various quantum communication and quantum cryptography
schemes~\cite{ReviewQC}.

\begin{acknowledgments}
The authors wish to thank Avi Pe'er for many fruitful discussions.
Financial support of this research by the Israel Science
Foundation is gratefully acknowledged.
\end{acknowledgments}

\end{document}